\documentstyle[aps,prl,twocolumn,epsfig]{revtex}
\begin{document}

\title{ Deconfinement Phase Transition in an Expanding Quark system
        in Relaxation Time Approximation }
\author{ Zhenwei Yang and Pengfei Zhuang\\
        Physics Department, Tsinghua University, Beijing 100084, China }
\maketitle

\begin{abstract}
\setlength{\baselineskip}{16pt} We investigated the effects of
nonequilibrium and collision terms on the deconfinement phase
transition of an expanding quark system in Friedberg-Lee model in
relaxation time approximation. By calculating the effective quark
potential, the critical temperature of the phase transition is
dominated by the mean field, while the collisions among quarks and
mesons change the time structure of the phase transition
significantly.
\end{abstract}
\noindent ${\bf PACS: 05.60.+w, 52.60.+h, 12.38.Mh}$

\section {Introduction}

It is generally accepted that the most interesting quantum
chromodynamics (QCD) phase transition in hot and dense nuclear
matter is the deconfinement phase transition between normal
nuclear matter and quark-gluon plasma (QGP), where quarks and
gluons are no longer confined. The theoretical and experimental
investigation of QGP is one of the most challenging problems in
high energy physics. It is widely believed that QGP phase can be
formed in ultra-relativistic heavy-ion collisions.

Due to the difficulty of non-perturbative treatment in QCD,
various models have been considered in the study of the phase
transition among which Friedberg-Lee model\cite{fl}, also referred
as nontopological soliton model, has been widely discussed in the
past two decades, see for instance \cite{lp,wilets} and refs.
therein. In this model, the non-perturbative dynamics responsible
for confinement in QCD is simulated in terms of a non-linear
coupling to a scalar field $\sigma$. It shows an intuitive
mechanism for the deconfinement phase transition. In vacuum state,
the physical value of $\sigma$ is large and the quark mass is more
than $1$ GeV, so that the effective heavy quarks have to be
confined in hadron bags\cite{fl}. With increasing temperature
and/or density of the system, the physical value of $\sigma$ and
in turn the effective quark mass drops down, the thermodynamic
motion leads to a deconfinement of the effective light quarks.

Wilets and his cooperators\cite{wilets,gw} did a great deal of
work on the properties of Friedberg-Lee model, mostly in mean
field approximation and in vacuum state. It is proved very
successful in describing the static properties of the nucleon.
During the past years, Friedberg-Lee model was extended to finite
temperature and density to study deconfinement phase
transition\cite{rds,lbw,wll,gwl}. Similar to most of these
investigations in the frame of finite temperature field theory,
the temperature and density effect on the phase transition is
based on the assumption of a thermalized plasma phase. While one
can use various parameters and take different treatments, the
critical temperature of the deconfinement phase transition in the
model is limited in the region of 80 to 120 MeV at zero chemical
potential, much lower than the prediction in Lattice
QCD\cite{karsch}.

Because of the estimated very short lifetime of the heavy ion
collision zone, the highly excited particle system may spend a
considerable fraction of its life in a nonthermalized,
preequilibrium state. The dynamical tool to treat dissipative
processes in heavy ion collisions and the approach to local
thermal equilibrium is in principle nonequilibrium quantum
transport theory. A relativistic and gauge covariant kinetic
theory for quarks and partons has been derived\cite{eh}, both in a
classical framework\cite{heinz1,heinz2} and as a quantum kinetic
theory\cite{egv1,egv2} based on Winger operators defined in
eight-dimensional phase space\cite{glw}. To solve the quantum
kinetic equations as initial problems, the transport and off-shell
constraint hierarchies have been established\cite{bgr,zh1,zh2,oh}
in the frame of equal-time Wigner operators. The properties of
nucleon-nucleon collisions have been explored\cite{kvbm,vbm} in
transport approach of Friedberg-Lee model.

In the present paper, we consider an expanding non-equilibrium
system with collision terms in the framework of Friedberg-Lee
model to investigate the approach of strong interaction matter
toward thermal equilibrium and the deconfinement phase transition
during this process. The collision terms will be introduced
through a relaxation time approximation in transport equations of
quarks and sigmas, and the expansion of the system is simply
described by the Bjorken scaling hydrodynamics\cite{bjorken}. We
specially focus on the effect of collision terms and
non-equilibrium on the deconfinement phase transition, by
investigating at what proper time the phase transition occurs, how
long it lasts for a first-order transition, and their dependence
on the relaxation time.

The outline of the paper is as follows. The full transport
equations for quarks and sigmas in relaxation time approximation,
and the simplified equations in quasiparticle limit and boost
invariance approximation are presented in Section 2. In section 3
we exhibit the numerical results and discussions. Finally a brief
summary is given in the last section.

\section {Transport Equations }

The Friedberg-Lee model is defined as\cite{fl,lp,wilets}
 \begin{eqnarray}
 \label{fl}
 {\cal L}_{FL} &=& \hat{\bar\psi}\left(i\gamma^\mu\partial_\mu-(m_0+g\hat\sigma)\right)\hat\psi
                   +{1\over 2}\partial^\mu\hat\sigma\partial_\mu\hat
                   \sigma-U(\hat\sigma),\nonumber\\
  U(\hat\sigma) &=& {a\over 2}\hat\sigma^2+{b\over 3!}\hat\sigma^3
                     +{c\over 4!}\hat\sigma^4+B\ ,
 \end{eqnarray}
where $\hat\psi, \hat{\bar\psi}$ and $\hat\sigma$ are quark,
antiquark and scalar fields, respectively, $m_0$ is the current
quark mass and chosen to be $0$ in the following to simplify the
calculations. There are five parameters in the Friedberg-Lee
model. $a$ with dimension $L^{-2}$, $b$ with dimension $L^{-1}$,
dimensionless $c$, the coupling constant $g$ between quark and
scalar fields, and the bag constant $B$ used to provide a quark
confinement potential in mean field approximation in the vacuum.
Of the five parameters, two are adjusted to fit the proton size
and the nucleon mass, and the others are left to survey, fit and
predict physical data. Since the model is an effective one, the
values of the parameters depend on the level of approximation
employed\cite{hl}. Different parameter sets can be found in the
book by Wilets\cite{wilets}.

The relativistically covariant quark Wigner operator $\hat W_q$
and sigma Wigner operator $\hat W_\sigma$ are the Fourier
transform of the corresponding density matrices\cite{glw}
\begin{eqnarray}\label{wo}
\hat W_q(x,p) &=& \int d^4 y e^{ipy}\hat\Phi_q(x,y)\nonumber\\
              &=& \int d^4y e^{ipy}\hat{\psi}(x+{y\over 2})
             \hat{\bar\psi}(x-{y\over 2})\ ,\nonumber\\
\hat W_\sigma(x,p) &=& \int d^4 y e^{ipy}\hat\Phi_\sigma(x,y)\nonumber\\
                   &=& \int d^4ye^{ipy} \hat\sigma(x+{y\over 2})\hat\sigma(x-{y\over 2})\ .
\end{eqnarray}

Calculating the first-order derivatives of $\hat\Phi_q$ and
second-order derivatives of $\hat\Phi_\sigma$ with respect to $x$
and $y$, and making use of the equations of motion
 \begin{eqnarray}
 \label{field}
  && \left(i\gamma^\mu\partial_\mu-(m_0+g\hat\sigma)\right)\hat\psi = 0\ ,\nonumber\\
  && \partial^\mu\partial_\mu\hat\sigma+{\partial U(\hat\sigma)\over
     \partial{\hat\sigma}}+g\hat{\bar\psi}\hat\psi = 0
 \end{eqnarray}
for the fields, one obtains evolution equations for the density
matrices. After Wigner transform one derives the kinetic equations
(for details, see the similar work in \cite{zh1} for QED)
\begin{eqnarray}
 \label{wf1}
 \left(K_\mu K^\mu - \Sigma_\sigma\right)W_\sigma(x,p) &=& C_\sigma\ ,\nonumber\\
 \left(\gamma^\mu K_\mu-\Sigma_q\right)W_q(x,p) &=& C_q
 \end{eqnarray}
for the Wigner functions $W_q(x,p)$ and $W_\sigma(x,p)$ which are,
respectively, the ensemble average of the Wigner operators,
\begin{eqnarray}\label{wf}
W_q(x,p) &=& \int d^4y e^{ipy}\langle\hat{\psi}(x+{y\over 2})
             \hat{\bar\psi}(x-{y\over 2})\rangle\ ,\nonumber\\
W_\sigma(x,p) &=& \int d^4ye^{ipy} \langle \hat\sigma(x+{y\over
2})\hat\sigma(x-{y\over 2})\rangle\ .
\end{eqnarray}
The operators $K_\mu, \Sigma_\sigma$ and $\Sigma_q$ are defined as
\begin{eqnarray}
 \label{selfenergy}
 K_\mu &=& p_\mu+\frac{i\hbar}{2}\partial_\mu\ , \nonumber \\
 \Sigma_\sigma &=& \left(a+b\sigma+{c\over 2}\left(\langle\hat\sigma'\hat
                   \sigma'\rangle+\sigma^2\right)\right) e^{-{i\over
                   2}\hbar\partial_x^\mu\partial^p_\mu}\ , \nonumber\\
 \Sigma_q &=& \left(m_0+g\sigma\cos\left({\hbar\over 2}\partial_x^\mu\partial^p_\mu\right)\right)
              - ig\sigma\sin\left({\hbar\over
              2}\partial_x^\mu\partial^p_\mu\right)\ ,
\end{eqnarray}
where $\sigma$ and $\hat\sigma'$ are, respectively, the mean field
and quantum fluctuation of the scalar field,
$\sigma=\langle\hat\sigma\rangle$ and $\hat\sigma =
\sigma+\hat\sigma'$. The sigma and quark scalar densities
$\langle\hat\sigma'\hat\sigma'\rangle$ and
$\langle\hat{\bar\psi}\hat\psi\rangle$ can be calculated through
the Wigner functions
 \begin{eqnarray}
 \label{density}
  && <\hat\sigma'\hat\sigma'> = \int{d^4p\over (2\pi)^4}W_\sigma(x,p)\ ,\nonumber\\
  && <\hat{\bar\psi}\hat\psi> = Tr\int{d^4p\over (2\pi)^4} W_q(x,p)\ .
 \end{eqnarray}
In relaxation time approximation the collision terms $C_\sigma$
and $C_q$ can be written as
\begin{eqnarray}
 \label{rta}
  C_\sigma&=&-i\hbar p^{\mu}u_{\mu}{{W_\sigma-W_{\sigma}{^{th}}}\over
  \theta},\nonumber\\
  C_q&=&-{i\hbar \over 2}\gamma^\mu u_\mu {{W_q-W_q{^{th}}}\over
  \theta}\ ,
\end{eqnarray}
where $u_\mu$ is the four-velocity of the hot medium formed by
quarks and sigmas, and $\theta$ is the relaxation time. It is
necessary to note that the space-time derivative $\partial_x$ in
the self-energies $\Sigma_\sigma$ and $\Sigma_q$ works only the
mean field $\sigma$ and scalar density
$\langle\hat\sigma'\hat\sigma'\rangle$ on its left.

Making ensemble average of the Klein-Gordon equation in
(\ref{field}), one gets the equation of motion for the condensate
$\sigma$,
 \begin{equation}
 \label{sigma}
  \partial^\mu\partial_\mu\sigma + {\partial U_{eff}\over \partial\sigma} = 0
 \end{equation}
with the definition of the effective confinement potential
$U_{eff}(\sigma, \langle \hat\sigma'\hat\sigma'\rangle,
\langle\hat{\bar\psi}\hat\psi\rangle)$ at finite temperature
\begin{eqnarray}\label{v1}
{\partial U_{eff}\over \partial\sigma} = \frac{\partial
U(\sigma)}{\partial\sigma} +
\frac{\langle\hat\sigma'\hat\sigma'\rangle}{2}(b+c\sigma)+g\langle\hat{\bar\psi}\hat\psi\rangle
\end{eqnarray}
which depends on not only the mean field but also the sigma and
quark scalar densities. The Eqs.(\ref{wf1}) and (\ref{sigma})
together determine the sigma and quark scalar distributions and
the condensate $\sigma$ self-consistently.

From the first-order derivative of the effective potential
$U_{eff}$, Eq.(\ref{v1}), it can be defined as
 \begin{eqnarray}
 \label{poten}
&& U_{eff}(\sigma, \langle \hat\sigma'\hat\sigma'\rangle,
\langle\hat{\bar\psi}\hat\psi\rangle)=\nonumber\\
&& \ \ \ \ \  \ \ \ \ \ \ \ U(\sigma) + U_\sigma(\sigma, \langle
\hat\sigma'\hat\sigma'\rangle) + U_q(\sigma,
\langle\hat{\bar\psi}\hat\psi\rangle),\nonumber\\
&& U_\sigma(\sigma, \langle \hat\sigma'\hat\sigma'\rangle)
  = \int_0^\sigma d\tilde\sigma\frac{1}{2}\langle\hat \sigma' \hat
                         \sigma'\rangle(\tilde\sigma,x)\left(b+c\tilde\sigma\right),\nonumber\\
&& U_q(\sigma, \langle\hat{\bar\psi}\hat\psi\rangle) =
\int_0^\sigma d\tilde\sigma g\langle\hat{\bar \psi}
\hat\psi\rangle(\tilde\sigma,x)\ .
\end{eqnarray}
The two extra terms $U_\sigma$ and $U_q$ arise from the collective
motion of the quarks and sigmas. In the vacuum without collective
motion the confinement potential $U_{eff} = U$ has two minima,
$\sigma_{per} = 0$ corresponding to the perturbative vacuum and
$\sigma_{phy}\ne 0$ to the physical vacuum, see the dashed line in
Fig.\ref{fig1}. From the definition of the bag constant, the
energy density difference between the perturbative and physical
vacua, $B$ is determined by $U(\sigma_{phy}) = 0$,
\begin{equation}
-B=\frac{a}{2!}\sigma_{phy}^2+\frac{b}{3!}\sigma_{phy}^3+\frac{c}{4!}\sigma_{phy}^4.
\end{equation}
The bag constant $B$ depends strongly on the parameters $a,\ b,\
c$ and $g$. For various reasonable sets of parameters, however,
$B$ is around 20 MeV/fm$^3$\cite{wilets}. As discussed in
\cite{rds}, it is the small bag constant $B$ which leads to the
low critical temperature of deconfinement phase transition in
Friedberg-Lee model.

An important aspect of the covariant kinetic theory is that the
complex kinetic equation can be split up into a constraint and a
transport equation\cite{egv1,egv2,zh1,zh2,oh}, where the former is
a quantum extension of the classical on-shell condition, and the
latter is a covariant generalization of the Vlasov-Boltzmann
equation. The complementarity of these two ingredients is
essential for a physical understanding of quantum kinetic theory.
The quark Wigner function itself has no direct physical analogue
since it is not a self-hermitian function. After the Lorentz
decomposition\cite{egv1,egv2,zh1,zh2,oh}, the kinetic equation for
quark is changed into 16 transport equations and 16 constraint
equations for the 16 Lorentz components of the quark Wigner
function. In classical limit with $\hbar=0$, the constraint
equations for quark and sigma are reduced to
\begin{eqnarray}
\label{onshell}
\left(p^2-m_\sigma^2\right)W_\sigma(x,p) &=& 0\ ,\nonumber\\
\left(p^2-m_q^2\right)W_q(x,p) &=& 0
\end{eqnarray}
with effective sigma mass and quark mass
 \begin{eqnarray}
 \label{mass}
 m_\sigma^2 &=& a+b\sigma+{c\over 2}(\langle\hat\sigma'\hat\sigma'\rangle+\sigma^2)\ ,
     \nonumber\\
 m_q &=& m_0+g\sigma\ ,
\end{eqnarray}
and the 16 spinor components are no longer fully independent. Only
the quark number density $f_q$ and the spin density $\vec g_0$ are
the fundamental elements and the other components can be expressed
in terms of $f_q$ and $\vec g_0$\cite{zh1,zh2}. The classical
transport equations for sigma and quark densities $f_\sigma$ and
$f_q$ are reduced to the familiar Boltzmann equations in the rest
frame of the heat bath where the four-velocity is $u_\mu =
\{1,\vec 0\}$,
 \begin{equation}
 \label{class}
   \partial_t f_{\sigma,q} +({{\bf p}{\cdot}{\bf \nabla}\over E_{\sigma,q}})
    f_{\sigma,q}-{{\bf \nabla}m_{\sigma,q}^2\over 2 E_{\sigma,q}}{\cdot}{\bf \nabla}_p
    f_{\sigma,q} = -{f_{\sigma,q}-f_{\sigma,q}^{th}\over\theta}
 \end{equation}
with the particle energies $E_{\sigma,q} =
\sqrt{m_{\sigma,q}^2+p^2}$. The relations between the scalar
densities and number densities are
 \begin{eqnarray}
 \label{relation}
  && \langle\hat\sigma' \hat\sigma'\rangle = \int{d^3{\bf p}\over (2\pi)^3}
     {1\over E_\sigma}f_\sigma(x,{\bf p})\ ,\nonumber\\
  && \langle\hat{\bar\psi}\hat\psi\rangle = m_q\int{d^3{\bf p}\over (2\pi)^3}
     {1\over E_q}f_q(x,{\bf p})\ .
 \end{eqnarray}
The equilibrium distribution functions in the classical transport
equations (\ref{class}) are the familiar Bose-Einstein and
Fermi-Dirac distributions
 \begin{equation}
 \label{thermal}
  f_{\sigma,q}^{th}(x,{\bf p}) = {g_{\sigma,q}\over e^{E_{\sigma,q}\over T}\mp 1}
 \end{equation}
with the sigma and quark degenerates $g_\sigma = 1$ and $g_q=24$.

The classical transport equations can be greatly simplified by
taking into account Bjorken's boost invariant picture often used
to describe longitudinal expansion of relativistic heavy ion
collisions where a central plateau of the final rapidity
distribution exists\cite{bjorken}. Baym extended Bjorken's method
of scalar hydrodynamics to the phase space and solved the
Vlasov-Boltzmann equation with a constant particle
mass\cite{baym}.

Neglecting transverse expansion of the system and assuming boost
invariance along the longitudinal direction make the transport
equations (\ref{class}) for sigma and quark with effective masses
much more simple, depending on the proper time $\tau =
\sqrt{t^2-z^2}$ only,
 \begin{equation}
 \label{boost1}
   \partial_\tau f_{\sigma,q} = -{f_{\sigma,q}-f_{\sigma,q}^{th}\over\theta}
 \end{equation}
with the solutions
 \begin{eqnarray}
 \label{boost2}
   f_{\sigma,q}(\tau,{\bf p}) &=& e^{-\int_{\tau_0}^\tau {d\tau'\over \theta(\tau')}}
                         f_{\sigma,q}(\tau_0,p_T,p_z\frac{\tau}{\tau_0})\nonumber\\
   &+&\int_{\tau_0}^{\tau}\frac{d\tau'}{\theta(\tau')}e^{-\int_{\tau'}^{\tau}{d\tau''\over
   \theta(\tau'')}}f_{\sigma,q}^{th}(T,p_T,p_z\frac{\tau}{\tau'})\ ,
 \end{eqnarray}
where the time ratios $\tau\over \tau_0$ and $\tau\over \tau'$
come from the longitudinal expansion of the system\cite{baym}.

While the quark distribution in (\ref{boost2}) is the same as that
obtained in Baym scenario\cite{baym}, the quark mass here depends
on the mean field $\sigma$. This dependence couples the quark
distribution $f_q$ with the meson distribution $f_\sigma$ and the
mean field $\sigma$. It is this coupling that leads to the phase
transition from confinement to deconfinement through the change of
the quark mass with the mean field. In the longitudinal boost
invariant picture, the equation (\ref{sigma}) of motion for the
scalar condensate $\sigma$ is simplified as a normal second-order
derivative equation,
 \begin{equation}
 \label{boost3}
  \left(\partial_\tau^2+{1\over \tau}\partial_\tau\right)\sigma
  +{\partial U\over \partial\sigma}+{1\over
  2}\langle\hat\sigma'\hat\sigma'\rangle
  (b+c\sigma)+g\langle\hat{\bar\psi}\hat\psi\rangle = 0\ .
 \end{equation}

The proper time dependence of the temperature, $T(\tau)$,  in the
equilibrium distribution $f^{th}$ is determined by the energy
conservation law in collisions,
\begin{eqnarray}
 \label{boost4}
&&  \epsilon(\tau,\sigma) =
    \epsilon^{th}(T,\sigma)\ ,\nonumber\\
&&  \epsilon(\tau,\sigma) = \int {d^3 \bf p \over
    (2\pi)^3}\left[E_\sigma f_\sigma(\tau,\sigma,\vec p)
    + E_q f_q(\tau,\sigma,\vec p)\right]\ ,\nonumber\\
&& \epsilon^{th}(T,\sigma) = \int {d^3 \bf p \over
   (2\pi)^3}\left[E_\sigma f_\sigma^{th}(T,\sigma,\vec p)
   +E_q f_q^{th}(T,\sigma,\vec p)\right]\ ,\nonumber\\
 \end{eqnarray}
where we kept only the particle contribution to the energy
density, and neglected the mean field terms in $\epsilon$ and
$\epsilon^{th}$, since they are the same and do not affect the
energy conservation law.

 In the relaxation time approximation, the key step is the
computation of the relaxation time $\theta$. In principle,
$\theta$ depends on the type of particles, sigma or quark, and is
a function of phase space coordinates. For all the discussions
above we have neglected its type dependence and momentum
dependence. Qualitatively, $\theta$ has the order of the standard
strong interaction scale, $\theta \sim 1 fm$\cite{baym}.
Considering the fact that collision terms are the driving force
for the system to reach equilibrium, they affect the system
strongly only in the beginning of the evolution and are damped
with increasing time. With this feature in mind we model the
relaxation time by a simple step function,
\begin{equation} \label{theta}
{1\over \theta(\tau)} = {1\over \lambda} \Theta(\tau_e-\tau)\ ,
\end{equation}
where $\lambda$ is an inverse measure of the collision strength
and $\tau_e$ indicates the collision duration scale. $\lambda$ has
the order of 1 fm and $\lambda\rightarrow\infty$ means equilibrium
without collision. In order to get a more realistic
parametrization we replace in the following numerical calculations
the $\Theta$ function by a more gradual exponential function
$\left(1+e^{\tau-\tau_e\over l}\right)^{-1}$ with the collision
duration variance $l$. When $l\rightarrow 0$, it goes back to the
step function.

In order to solve numerically the transport equations for the
distributions $f_{\sigma,q}$ and the Klein-Gordon equation for the
mean field $\sigma$ as functions of proper time $\tau$, one must
know their initial values. Since the particles produced in
relativistic heavy ion collisions at initial time $\tau_0$ are
essentially emitted from the colliding point $z=0$ at $t=0$, the
initial distributions in the central slice are peaked in the plane
$p_z = 0$\cite{baym}. Therefore, we can choose the initial
distributions as
\begin{equation}\label{initial}
f_{\sigma,q}(\tau_0,\vec p) =
f_{\sigma,q}^{th}(T_0,\vec p)\delta (p_z)\ .
\end{equation}

With the known number distributions (\ref{boost2}) one can get the
scalar condensates $\langle\hat\sigma'\hat\sigma'\rangle$ and
$\langle\hat{\bar\psi}\hat\psi\rangle$ through the relations
(\ref{relation}) and in turn the effective quark potential
$U_{eff}$.

As we discussed above, there are two minima of the potential in
the vacuum without collective motion, $\sigma_{phy}$ is the global
minimum but $\sigma_{per}$ is only the local minimum, see the
dashed line in Fig.\ref{fig1}. This means that the system is in
physical vacuum. Since we focus in this paper on the phase
transition from deconfinement to confinement, which can be
realized in relativistic heavy ion collisions, we put initially
the system in a deconfinement state with strong enough collective
motion, and then study when the phase transition happens. To this
end, the temperature $T_0$ in the initial distributions
(\ref{initial}) should be high enough to guarantee that the system
is in perturbative vacuum initially, namely $\sigma_{per}$ is the
absolute minimum of the effective potential in the beginning. Due
to the expansion and the collision terms, the energy density and
effective temperature of the system fall down, and the potential
difference between the physical and perturbative vacua decreases
during the evolution. At some critical time $\tau_c$ the
difference between the two vacua disappears and a first-order
phase transition begins.

We define $\tau_h$ as the terminating time of the first-order
phase transition. During the process of the phase transition, the
deconfinement state and confinement state coexist, the energy
densities can be expressed as
\begin{eqnarray}
 \epsilon(\tau,\sigma)&=& x(\tau)\epsilon(\tau,\sigma_{per})+(1-x(\tau))
                          \epsilon(\tau,\sigma_{phy}),\nonumber\\
 \epsilon^{th}(T_c,\sigma)&=& x(\tau)\epsilon^{th}(T_c,\sigma_{per})+(1-x(\tau))
                              \epsilon^{th}(T_c,\sigma_{phy}),\nonumber\\
\end{eqnarray}
where $x(\tau)$ is the fraction of matter in the deconfinement
phase at time $\tau_c < \tau <\tau_h$. From the energy
conservation,
\begin{equation}\label{conse}
\epsilon(\tau,\sigma)=\epsilon^{th}(T_c,\sigma)\ ,
\end{equation}
one reads
\begin{eqnarray}
&&\label{x} x(\tau)=\nonumber\\
&&\frac{\epsilon^{th}(T_c,\sigma_{phy})-\epsilon(\tau,\sigma_{phy})}
             {\left[\epsilon(\tau,\sigma_{per})-\epsilon^{th}(T_c,\sigma_{per})\right]
             +\left[\epsilon^{th}(T_c,\sigma_{phy})-\epsilon(\tau,\sigma_{phy})\right]}\ .
             \nonumber\\
\end{eqnarray}
Clearly, at the beginning time $\tau_c$ and the end time $\tau_h$
of the first-order transition, one has
\begin{eqnarray}
 x(\tau_c) &=& 1\ ,\nonumber\\
 x(\tau_h) &=& 0\ .
\end{eqnarray}

\section { Numerical Results and Discussions }

We choose the potential parameters\cite{wilets,gw,wll} $ a=51.6
fm^{-2}, b=-799.9 fm^{-1}, c=4000$ and the coupling constant
$g=14.8$, which are satisfactory for fitting a part of the static
properties of hadrons and lead to a small bag constant $B= 28
MeV/fm^3$. The temperature in the initial distributions
(\ref{initial}) is taken as $T_0 = 160$ MeV which is high enough
to guarantee an initial deconfinement state. As explained above,
the collision parameter $\lambda$ describes the intensity of
collisions and therefore determines how long it needs to approach
the equilibrium state, and $\tau_e$ is the duration of the
collisions which controls how long the interaction lasts. As
discussed by Baym\cite{baym}, $\lambda\leq 1fm$ is reasonable for
strong interaction. According to the space-time scale of
relativistic heavy-ion collisions, one can estimate $\tau_e \sim
r_0 A^{1/3}$ fm.

\begin{figure}[ht]
\vspace*{+0cm} \centerline{\epsfxsize=7cm \epsffile{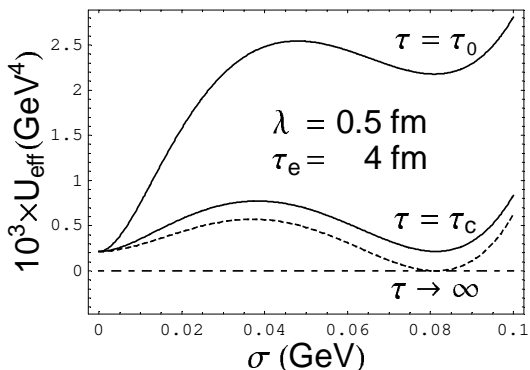}}
\caption{\it The effective potential $U_{eff}$ as a function of
$\sigma$ at initial time $\tau_0$, critical time $\tau_c$, and the
limit $\tau\rightarrow\infty$. The deconfinement phase transition
begins at $\tau_c = 5.6 fm$ and ends at $\tau_h = 30 fm$.}
\label{fig1}
\end{figure}

Fig.\ref{fig1} shows the effective potential $U_{eff}$ as a
function of $\sigma$ at different time for collision parameters
$\lambda=0.5$ fm and $\tau_e=4$ fm. At initial time $\tau_0$,
$U_{eff}$ has two minima, the local one at $\sigma=\sigma_{phy}$
and the global one at $\sigma=\sigma_{per}=0$. The system stays
initially in the perturbative vacuum. As time goes on, the system
expands and its energy density drops down gradually. While the
positions of the two minima remain unchanged during the evolution,
the potential difference between the two minima, the effective bag
constant $B_{eff}(\tau)$, becomes smaller and smaller. At a
critical time $\tau_c$, $B_{eff}(\tau_c)=0$, the first-order phase
transition begins and the confinement phase appears. After
$\tau_c$, the temperature parameter in $f_{\sigma,q}^{th}(T,\vec
p)$ remains a constant $T_c$, the physical vacuum and the
perturbative vacuum coexist, namely the deconfinement phase and
confinement phase coexist until another critical time $\tau_h$
when the transition is totally completed and the system is purely
in confinement state. For the collision parameters used, the
numerical calculation gives $\tau_c=5.6$ fm, $\tau_h=30.4$ fm, and
$T_c = 94$ MeV. After $\tau_h$, the temperature parameter drops
down again, and the global minimum is located at
$\sigma=\sigma_{phy}$. When $\tau\rightarrow\infty$, the effective
potential $U_{eff}$ approaches to $U$ in the vacuum, shown as the
dashed line in Fig.{\ref{fig1}. Fig.\ref{fig2} shows the evolution
of the temperature parameter before $\tau_c$.

\begin{figure}[ht]
\vspace*{+0cm} \centerline{\epsfxsize=7cm \epsffile{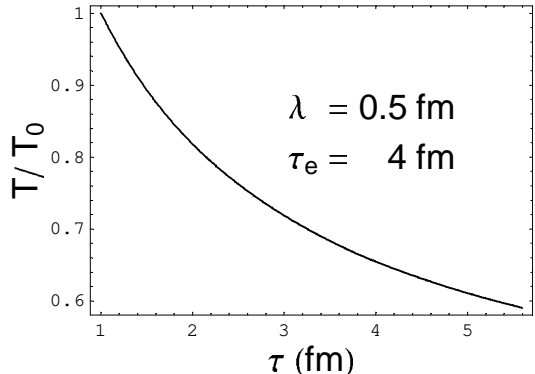}}
\caption{\it The temperature parameter as a function of proper
time before the phase transition. } \label{fig2}
\end{figure}

The two critical times, $\tau_c$ and $\tau_h$, strongly depend on
the relaxation time parameters, $\lambda$ and $\tau_e$. A system
with strong interaction, namely small $\lambda$, needs only a
short time to approach equilibrium state, and the temperature
parameter drops down rapidly. Correspondingly, the phase
transition occurs early and the duration of the first-order
transition is short. The parameter $\tau_e$ controls also the
values of $\tau_c$ and $\tau_h$. A large $\tau_e$ means a long
last of collisions, the system needs only a short time to get to
the points where the phase transition begins and ends. If the
duration of the collision $\tau_e$ is long enough, the system will
approach equilibrium state eventually. If $\tau_e$ is too small,
however, the system will not approach equilibrium state unless the
collision strength is strong enough. Fig.\ref{fig3} shows the
effective potential as a function of $\sigma$ in the limit of
$\lambda\rightarrow\infty$. In this case, the only driving force
of the phase transition is the cooling of the mean field system,
the dynamics is purely reflected in the effective particle masses
$m_\sigma$ and $m_q$. The transport equations (\ref{boost2}) are
reduced to
\begin{equation}
 \label{sln1tt}
  f_{\sigma,q}(\tau,{\bf p}) = f_{\sigma,q}(\tau_0,\vec p_T,p_z{\tau\over \tau_0})\ .
 \end{equation}
If we neglect the collective motion of $\sigma$ since it is heavy
enough, and consider the fact that before the phase transition
quarks are massless in the deconfinement state, the energy
conservation law in the limit of $\lambda\rightarrow\infty$ gives
approximately the time evolution of the temperature parameter,
$T(\tau) \sim T_0 \left(\tau_0/\tau\right)^{1/4}$. Only when the
thermal distribution in (\ref{initial}) is taken as Boltzmann
distribution, we have exactly,
\begin{equation}\label{temp1}
T(\tau)=T_0 \left({\tau_0\over\tau}\right)^{1/4}\ .
\end{equation}

If we put the initial state in a 3-dimensional equilibrium state
and take the mass to be zero, the longitudinal expansion of the
system will force the state in nonequilibrium afterwards. The
energy conservation in the limit of no collisions leads to
\begin{eqnarray}\label{temp2}
T(\tau) &=& T_0 \left({\tau_0\over\tau}\right)^{1/4}g(\tau)\
,\nonumber\\
g(\tau) &=& \left[{1\over
2}\left({\tau_0\over\tau}+{\sin^{-1}\sqrt{1-\left({\tau_0\over\tau}\right)^2}\over
\sqrt{1-\left({\tau_0\over\tau}\right)^2}}\right)\right]^{1\over
4}\ .
\end{eqnarray}
Since the quark mass is taken as a constant, $m=0$ in this special
case, the quark distribution is decoupled from the meson
distribution and the mean field, the system is reduced to the case
in Baym scenario. Therefore, the result (\ref{temp2}) has a
similar structure to eq. (19) of the work by Baym\cite{baym}.

The comparison between (\ref{temp1}) and (\ref{temp2}) is shown in
Fig.\ref{fig4}. Since the function $g(\tau) \leq 1$, the
temperature with initial equilibrium state is always lower than
that with initial nonequilibrium state for any time $\tau >
\tau_0$.

\begin{figure}[ht]
\vspace*{+0cm} \centerline{\epsfxsize=7cm \epsffile{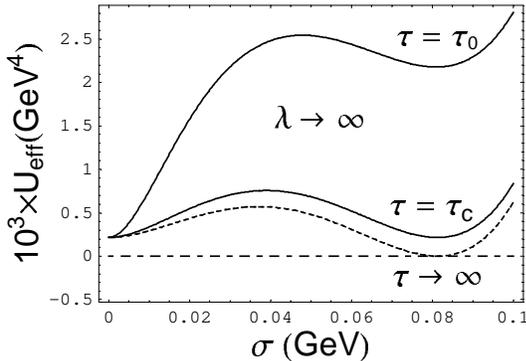}}
\caption{\it The effective potential $U_{eff}$ as a function of
$\sigma$ at initial time $\tau_0$, critical time $\tau_c$, and the
limit $\tau\rightarrow\infty$ in the case without collisions. In
this limit, the phase transition starts at $\tau_c = 10 fm$ and
ends at $\tau_h = 841 fm$.} \label{fig3}
\end{figure}

\begin{figure}[ht]
\vspace*{+0cm} \centerline{\epsfxsize=7cm \epsffile{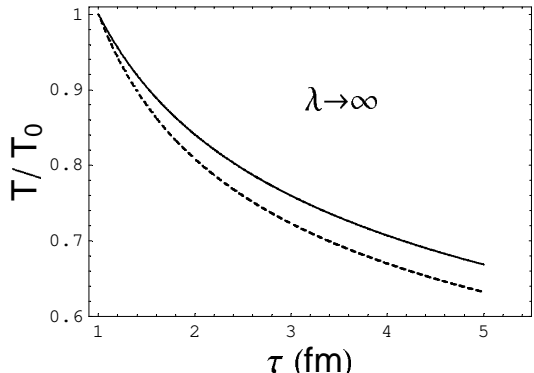}}
\caption{\it The time evolution of the temperature parameter with
initial nonequilibrium state (solid line) and with initial
equilibrium state (dashed line) in the limit of no collisions. }
\label{fig4}
\end{figure}

While the collisions do not change the critical value of the
temperature parameter remarkably, it is $94$ MeV for $\lambda =
0.5$ fm and $\tau_e = 4$ fm and $89$ MeV for
$\lambda\rightarrow\infty$, the time structure of the phase
transition is controlled by the collisions. The relaxation time
dependence of the critical time $\tau_c$ and the duration
$\Delta\tau=\tau_h-\tau_c$ scaled by their values in the limit of
no collisions is shown in Figs.\ref{fig5} and \ref{fig6}. Both
decrease monotonously with increasing collision strength and/or
increasing collision duration. When the collisions are strong
enough and the collision duration is long enough, the phase
transition will start before the collisions cease. This is clearly
reflected in Fig.\ref{fig5} where $\tau_c$ at $\lambda = 0.2$ fm
no more changes when $\tau_e > 4$ fm. With increasing $\lambda$
the starting time $\tau_c$ and the duration $\Delta\tau$ will
approach finally the limit values of no collisions, they converge
at $10.12$ fm and $831$ fm, respectively. The speed of the
convergence is governed by the collision duration. When the
collision duration is short enough, the system is not sensitive to
the collisions in the region of weak strength. This is the reason
why $\tau_c$ and $\Delta \tau$ converge very fast with small
$\tau_e$.

\begin{figure}[ht]
\vspace*{+0cm} \centerline{\epsfxsize=7cm \epsffile{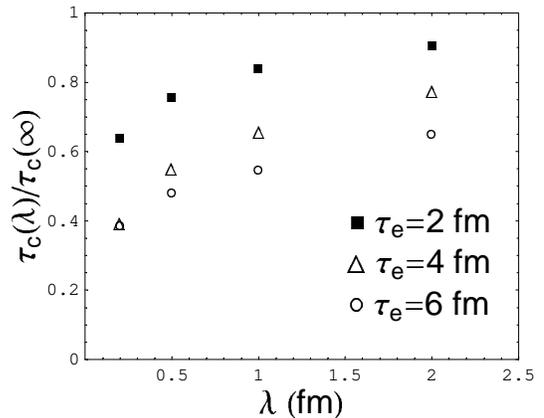}}
\caption{\it The beginning time of first-order phase transition
scaled by its limit value without collisions as a function of the
collision strength for collision duration $\tau_e = 2, 4, 6$ fm. }
\label{fig5}
\end{figure}

\begin{figure}[ht]
\vspace*{+0cm} \centerline{\epsfxsize=7cm \epsffile{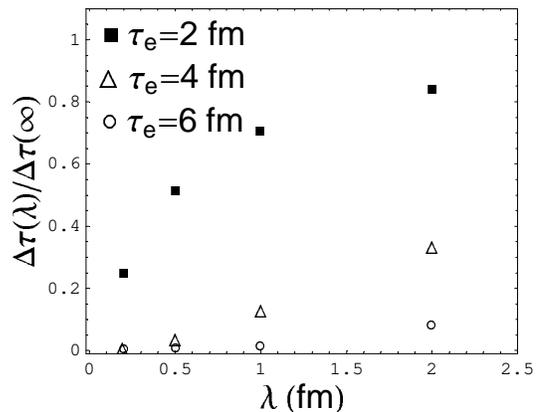}}
\caption{\it The duration of the first-order phase transition
scaled by its limit value without collisions as a function of the
strength for collision duration $\tau_e = 2, 4, 6$ fm. }
\label{fig6}
\end{figure}

Fig.\ref{fig7} shows the temperature parameter as a function of
time for fixed collision duration $\tau_e=4$ fm but different
values of collision strength. Each line starts at initial time
$\tau_0 = 1$ fm and ends at the critical time $\tau_c$ which
depends on the collision strength. After $\tau_c$ the temperature
does not change until the second critical time $\tau_h$. With
increasing collision strength, namely decreasing $\lambda$, the
system cools down more and more fast, and the phase transition
happens more and more early.

In principle, only one of the two collision parameters $\lambda$
and $\tau_e$ is free. A system with strong interaction needs a
short time to approach equilibrium, and vice versa. While it is
difficult to determine the relation between the two parameters in
relaxation time approximation, we can obtain a constraint
condition. For a system with fixed collision duration scale
$\tau_e$, the collision strength should larger than the maximum
value corresponding to the well-known Bjorken scaling solution
$T/T_0 = \left(\tau_0/\tau\right)^{1/3}$\cite{bjorken}. Since
$\lambda$ is the inverse measure of the collision strength, we
have the constraint
\begin{equation}
\label{constraint} \lambda_{min} < \lambda \ .
\end{equation}
For $\tau_e = 4 fm$, $\lambda_{min}$ is approximately $0.3 fm$.
When $\lambda$ is less than $\lambda_{min}$, the result will not
be physical.

\begin{figure}[ht]
\vspace*{+0cm} \centerline{\epsfxsize=7cm \epsffile{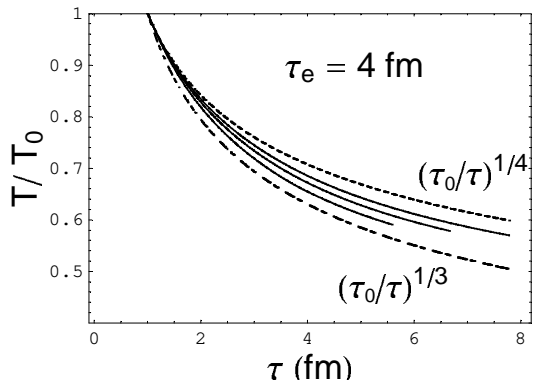}}
\caption{\it The temperature parameter $T$ as a function of proper
time $\tau$ at fixed $\tau_e$. The solid curves from the bottom to
the top correspond to $\lambda = 0.5, 1$ and $2$ fm, and the two
dashed lines correspond to the limit (\ref{temp1}) and the Bjorken
limit, respectively.} \label{fig7}
\end{figure}

\section { Summary }

We investigated the deconfinement phase transition of an expanding
quark system in Friedberg-Lee model, considering the collision
terms in relaxation time approximation. We calculated the
beginning time and the duration of the first-order phase
transition for different collision strength. While the critical
temperature of the phase transition is dominated by mean field,
described by $\lambda\rightarrow\infty$ in our treatment, the
collision terms have significant influence on the beginning time
and the duration of the transition. Strong collisions result in an
early and short transition, and weak collisions make the
transition begin late and last a long time. Although our results
are derived in a particular model, we expect the qualitative
dependence of the phase transition times on the collisions to be
of more general validity, because it is well known that collisions
are the driving force of thermalization and control the speed of
thermalization process of any system.

Since the deconfinement phase is only an intermediate state in
relativistic heavy ion collisions, there is no way to measure its
properties directly. One needs to extract signatures of the phase
transition and the new state from analysis of final state
distributions of different particles, such as low moment dilepton
enhancement, $J/\Psi$ suppression and strangeness enhancement at
SPS\cite{sps} and disappearance of back-to-back jets at
RHIC\cite{rhic}. Normally the theoretical study on these
signatures is based on the assumption of equilibrium system.
According to our discussion above, the thermalization process due
to collisions among the particles make the phase transition happen
early and shorten the duration of the first-order phase
transition. As a consequence, the contribution from the
deconfinement phase and the coexisting phase to the final state
distributions will be changed, especially for those signatures
extracted from the early evolution.

{\bf Acknowledgments:} We thank Larry Mclerran who drew our
attention to the time dependence of collision terms. We are
grateful to Jisheng Chen, Jinfeng Liao and Xianglei Zhu for many
discussions during the work. The work was supported in part by the
grants NSFC19925519, 10135030, and G2000077407.

\end{document}